\documentclass[10pt,a4paper]{article}
\usepackage{amsmath}
\usepackage{amssymb}
\usepackage{amsfonts}
\usepackage{amstext}
\usepackage{amsbsy}
\usepackage[mathscr]{eucal}
\usepackage{graphicx}
\usepackage{color}
\newcommand{\beq}{\begin{equation}}
\newcommand{\eeq}{\end{equation}}
\newcommand{\beqa}{\begin{eqnarray}}
\newcommand{\eeqa}{\end{eqnarray}}
\newcommand{\ket}[1]{| #1 \rangle}

\title{\Large\textbf{Multipartite quantum systems and symplectic toric manifolds}}

\author{\textit{ Hoshang Heydari}\\
        \small\textit{Physics Department, Stockholm university 10691 Stockholm Sweden}\\
\\\small\textit{Email: hoshang@fysik.su.se}}
\begin{document}
\maketitle

\begin{abstract}
 In this paper we  study the geometrical structures of  multi-qubit
 states based on symplectic toric manifolds.  After a short review of
 symplectic toric manifolds, we discuss the space of a single quantum
 state in terms of these manifolds.  We also investigate entangled
 multipartite states based on moment map and  Delzant's construction
 of toric manifolds and algebraic toric varieties.
\end{abstract}


\maketitle

\section{Introduction}
\label{intro}

During recent years the geometrical, topological, and combinatorial structures of multipartite quantum systems have been parts of  research in the fields of foundations of quantum theory, quantum information, and quantum computing \cite{Miyake,Briand2,Levay1,Hosh4}. These mathematical methods are also  very important in solving complex problems and visualizing the difficult physical concepts in other branches of physics such as general relativity, gauge theory, and string theory \cite{Vafa}.

Recently we have investigated the combinatorial and geometrical structures of quantum systems using complex projective toric varieties \cite{Hosh1,Hosh2}.
In this paper we will establish a relation between  multipartite quantum states and
symplectic toric manifolds. These manifolds of dimension $2n$ are compact
connected symplectic manifolds
which have  effective  hamiltonian actions of $n$-torus with corresponding
 moment maps. In particular, in section \ref{variety} we give a short introduction
  to complex projective varieties. In section \ref{sec2} we will review the
  construction of symplectic toric manifolds. The construction is abstract
   but we  simplify our review in such way that this text becomes
   suitable for the readers with fair amount of knowledged in symplectic and
   differential geometry. In section \ref{sec3} we will study the structure
   of a single quantum system based on symplectic toric manifolds. Finally,
   in section \ref{sec4} we will in detail investigate the structures of
    composite  quantum systems based on Delzant's construction of toric
    manifolds and algebraic toric varieties. For readers who are interested
    in symplectic geometry and topology, we recommend the following books
    \cite{Guillemin1,Audin,Mcduff,Guillemin2,Silva1, Silva2}. The aim of the
    material presented in this paper is twofold. Firstly it can be serves as
    a short introduction to the symplectic toric manifolds for physicist  and
     secondly it includes new possible applications of these mathematical methods
      in the field of quantum information processing.
Through this paper we will use  the following notation
\begin{equation}\ket{\Psi}=\sum^{1}_{x_{m}=0}\sum^{1}_{x_{m-1}=0}\cdots
\sum^{1}_{
x_{1}=0}\alpha_{x_{m}x_{m-1}\cdots x_{1}}\ket{x_{m}x_{m-1}\cdots
x_{1}},
\end{equation}
 with $\ket{x_{m}x_{m-1}\cdots
x_{1}}=\ket{x_{m}}\otimes\ket{x_{m-1}}\otimes\cdots\otimes\ket{x_{1}}\in
\mathcal{H}_{\mathcal{Q}}=\mathcal{H}_{\mathcal{Q}_{1}}\otimes
\mathcal{H}_{\mathcal{Q}_{2}}\otimes\cdots\otimes\mathcal{H}_{\mathcal{Q}_{m}}
$ for a pure multi-qubit state, where $\mathcal{H}_{\mathcal{Q}}$ is the Hilbert space of a composite quantum system and $\mathcal{H}_{\mathcal{Q}_{j}}$, for all $j=1,2,\ldots,m$ are the Hilbert spaces of quantum subsystems.

\section{Complex projective variety }\label{variety}
In this short section we will give an introduction to the complex
affine and projective varieties
 \cite{Griff78,Mum76,Hart77}.
An affine $n$-space over a complex algebraic field $\mathbb{C}$
denoted $\mathbb{C}^{n}$ is the set of all $n$-tuples of elements of
$\mathbb{C}$. An element $P\in\mathbb{C}^{n}$ is called a point of
$\mathbb{C}^{n}$ and if $P=(a_{1},a_{2},\ldots,a_{n})$ with
$a_{j}\in\mathbb{C}$, then $a_{j}$ is called the coordinates of $P$.
 A Zariski
closed
 set in $\mathbb{C}^{n}$ is a set of common zeros of a finite number of polynomials
 from the polynomial
algebra $\mathbb{C}[z]=\mathbb{C}[z_{1},z_{2}, \ldots,z_{n}]$  in
$n$  variables with complex coefficients and the complement of a
 Zariski closed set is called a Zariski open set.
Given a set of $q$ polynomials $\{g_{1},g_{2},\ldots,g_{q}\}$ with
$g_{i}\in \mathbb{C}[z]$, we define a complex affine variety as
\begin{eqnarray}
&&\mathcal{V}(g_{1},g_{2},\ldots,g_{q})=\{P\in\mathbb{C}^{n}:
g_{i}(P)=0,~\text{for all}~1\leq i\leq q\}.
\end{eqnarray}
Let $\mathcal{V}$ be complex affine algebraic variety. Then an ideal
of $\mathbb{C}[z_{1},z_{2}, \ldots,z_{n}]$ is defined by
\begin{eqnarray}
&&\mathcal{I}(\mathcal{V})=\{g\in\mathbb{C}[z_{1},z_{2},
\ldots,z_{n}]: g(z)=0,~\text{for all}~z\in\mathcal{V}\}.
\end{eqnarray}
We can also define a coordinate ring of an affine variety
$\mathcal{V}$
 by
 \begin{equation}
\mathbb{ C}[\mathcal{V}]=\mathbb{C}[z_{1},z_{2},
 \ldots,z_{n}]/\mathcal{I}(\mathcal{V}).
 \end{equation}
 A complex projective space $\mathbb{P}^{n}$ is
defined to be the set of lines through the origin in
$\mathbb{C}^{n+1}$, that is,
\begin{equation}
\mathbb{P}^{n}=\frac{\mathbb{C}^{n+1}-\{0\}}{
(x_{1},\ldots,x_{n+1})\sim(y_{1},\ldots,y_{n+1})},~\lambda\in
\mathbb{C}-0,~y_{i}=\lambda x_{i}
\end{equation}
 for all $0\leq i\leq n+1 $. For
example $\mathbb{P}^{1}=\mathbb{C}\cup\{\infty\}$,
$\mathbb{P}^{2}=\mathbb{C}^{2}\cup\mathbb{P}^{1}=\mathbb{C}^{2}\cup\mathbb{C}\cup\{\infty\}$
and in general we have
$\mathbb{P}^{n}=\mathbb{C}^{n}\cup\mathbb{P}^{n-1}$. Given a set of
homogeneous polynomials $\{h_{1},h_{2},\ldots,h_{q}\}$ with
$h_{i}\in \mathbb{C}[z]$, we define a complex projective variety as
\begin{equation}
\mathcal{V}(h_{1},\ldots,h_{q})=\{O\in\mathbb{P}^{n}:
h_{i}(O)=0~\forall~1\leq i\leq q\},
\end{equation}
where $O=[a_{1},a_{2},\ldots,a_{n+1}]$ denotes the equivalent class
of point $\{a_{1},a_{2},\ldots,$ $a_{n+1}\}\in\mathbb{C}^{n+1}$. The
ideal and coordinate ring of complex projective variety can be
defined in similar way as in the case of complex algebraic affine
variety by considering the complex projective space and it's
homogeneous coordinate. The Zariski topology on $\mathbb{C}^{n}$ is
defined to be the topology whose  closed sets are the set
\begin{equation}
\mathcal{V}(\mathcal{I})=\{z\in \mathbb{C}^{n}:g(z)=0,~\text{for all}~g\in\mathcal{I}\},
\end{equation}
where $I\subset\mathbb{C}[z_{1},z_{2},
 \ldots,z_{n}]$ is an ideal.
Let $\mathcal{A}$ be a finitely generated $\mathbb{C}$-algebra without zero divisors.
 Then an $\mathcal{I}$ in $\mathcal{A}$ is called prime ideal if for $f,g\in\mathcal{A}$
  and $f,g\in \mathcal{I}$ implies that $f\in \mathcal{I}$ or $g\in \mathcal{I}$.
   An ideal $\mathcal{I}\subset\mathcal{A}$ is called maximal if $\mathcal{I}\neq\mathcal{A}$
   and the only proper ideal in $\mathcal{A}$ containing $I$ is $I$. The spectrum of the
   algebra $\mathcal{A}$ is the set
   \begin{equation}\mathrm{Spec} \mathcal{A}=\{\mathrm{prime}~ \mathrm{ideal}~
   \mathrm{in} ~\mathcal{A}\}
\end{equation}
 equipped with the Zariski topology and the maximal spectrum of
    \begin{equation}
    \mathrm{Specm} \mathcal{A}=\{\mathrm{maximal}~ \mathrm{ideal}~ \mathrm{in}~ \mathcal{A}\}
 \end{equation}
     equipped with the Zariski topology.  Now,
 let  $\mathcal{X}$ be an affine variety in $\mathbb{C}^{n}$ defined by polynomials
  $g_{1},g_{2},\ldots, g_{q}$ from the polynomial ring $\mathbb{C}[z_{1},z_{2},
 \ldots,z_{n}]$ and $\mathcal{I}=\langle g_{1},g_{2},\ldots, g_{q}\rangle$.
 Then we have $\mathcal{X}\simeq\mathrm{Specm} \mathbb{C}[\mathcal{X}]$.
 If $\mathcal{A}$ is finitely generated $\mathbb{C}$-algebra without zero
 divisors, then we called $\mathcal{X}_{\mathcal{A}}\simeq\mathrm{Specm} \mathbb{C}[\mathcal{X}]$ an abstract affine variety. Note that in this construction maximal ideals are points
  in $\mathcal{X}_{\mathcal{A}}$ and prime ideals are irreducible subvarieties.

Next, we will review the construction of the  Segre variety
\cite{Hosh4} for general multi-projective complex space.  Let
$(\alpha^{i}_{1},\alpha^{i}_{2})$  be points defined on the complex
projective space $\mathbb{P}^{2_{i}-1}$. Then the Segre map
\begin{equation}
\begin{array}{ccc}
  \mathcal{S}:\underbrace{\mathbb{P}^{1}\times\mathbb{P}^{1}
\times\cdots\times\mathbb{P}^{1}}_{m~\text{times}}&\longrightarrow&
\mathbb{P}^{2^{m}-1}
\end{array}
\end{equation}
is defined by $
 ((\alpha^{1}_{0},\alpha^{1}_{1}),\ldots,
 (\alpha^{m}_{0},\alpha^{m}_{1}))  \longmapsto
 (\alpha^{1}_{i_{1}}\alpha^{2}_{i_{2}}\cdots\alpha^{m}_{i_{m}})$, where $0\leq i_{m}\leq 1$.  The image of the Segre embedding
$\mathrm{Im}(\mathcal{S})$, which again is an intersection of
families of  hypersurfaces in $\mathbb{P}^{2^{m}-1}$, is called
Segre variety and it is given by
\begin{eqnarray}\label{eq: submeasure}
\mathrm{Im}(\mathcal{S})&=&\bigcap_{\forall
j}\mathcal{V}(\alpha_{x_{m}x_{m-1}\ldots
x_{1}}\alpha_{y_{m}y_{m-1}\ldots y_{1}} \\\nonumber&&-
\alpha_{x_{m}x_{m-1}\ldots x_{j+1}y_{j}x_{j-1}\ldots
x_{1}}\alpha_{y_{m}y_{m-1} \ldots y_{j+1} x_{j}y_{j-1}\ldots
y_{1}}),
\end{eqnarray}
where $\alpha_{i_{m}i_{m-1}\ldots i_{1}}$ with $i=x,y$ are
homogeneous coordinate on $\mathbb{P}^{2^{m}-1}$.  We will also
discuss the multi-projective Segre variety in the following
sections.

\section{Symplectic toric manifolds}\label{sec2}
 In this section we will give a short introduction to the symplectic manifolds but for a detail construction of these manifolds we recommend the following books \cite{Guillemin2,Silva1, Silva2}. Let $\mathcal{V}$ be a finite-dimensional vector space and
$\Omega\in \mathcal{V}\times \mathcal{V}\longrightarrow \mathbb{R}$ be a bilinear form on $\mathcal{V}$. Then $\Omega$ is called nondegenerate if $\Omega(e_{1},e_{2})=0$ for all $e_{2}\in \mathcal{V}$ implies that $e_{1}=0$. A symplectic form
on a vector space $\mathcal{V}$ is  a nondegenrate two-form $\Omega\in\Omega^{2}(\mathcal{V})$. Moreover, the pair $(\mathcal{V},\Omega)$ is called a symplectic vector space. If we suppose that $(\mathcal{V},\Omega)$ and $(\mathcal{V}^{'},\Omega^{'})$ are two symplectic vector spaces, then, a symplectomorphism $\phi$ between these vector spaces is a linear isomorphism $\phi:\mathcal{V}\longrightarrow\mathcal{V}^{'}$ such that $\Omega=\phi^{*}\Omega^{'}$, where $\phi^{*}\Omega^{'}(u,v)=\Omega^{'}(\phi(u),\phi(v))$. In case symplectomorphism $\phi$ exists, then $(\mathcal{V},\Omega)$ and $(\mathcal{V}^{'},\Omega^{'})$ are called symplectomorphic which is an equivalence relation if the vector spaces are even-dimensional.

A symplectic form on a manifold $\mathcal{M}$ is a closed two-form  which is nondegenerate at every point of $\mathcal{M}$. In analogy with symplectic vector space, a symplectic manifold is the pair $(\mathcal{M},\Omega)$. For a given $(z_{1},z_{2},\ldots,z_{n})\in \mathcal{M}=\mathbb{C}^{n}\simeq\mathbb{R}^{2n}$ the form
\begin{equation}
\Omega=\frac{i}{2}\sum^{n}_{i=1}dz_{i}\wedge d\overline{z}_{i}
\end{equation}
is a symplectic form on $\mathbb{C}^{n}$. If $(\mathcal{M},\Omega)$ is a symplectic manifold of dimension $2n$, then a Darboux chart for  $\mathcal{M}$ is a chart $(\mathcal{C},x_{1},x_{2},\ldots,x_{n},y_{1},y_{2},\ldots,y_{n})$ such that
$\Omega|_{\mathcal{C}}=\sum^{n}_{i=1}dx_{i}\wedge dy_{i}$.
Now, if  $X$ is a vector field on $\mathcal{M}$, then the map
\begin{equation}
\imath_{X}:\Omega^{p}(\mathcal{M})\longrightarrow\Omega^{p-1}(\mathcal{M})
\end{equation}
defined by $\imath_{X}\Omega(X_{1},X_{2},\ldots,X_{p-1})=\Omega(X,X_{1},X_{2},\ldots,X_{p-1})$, for any vector fields $X_{1},X_{2},\ldots,X_{p-1}$ is called interior product or contraction of differential form with a vector field.
A vector field  $X$ on $\mathcal{M}$ is symplectic if the contraction  $\imath_{X}\Omega$ is closed and is hamiltoninan if the contraction is exact. Moreover, a hamiltonian function for a hamiltonian vector field $X$ on $\mathcal{M}$ is a smooth function $H:\mathcal{M}\longrightarrow \mathbb{R}$ such that $\imath_{X}\Omega=dH$ and a hamiltonian system is a triple $(\mathcal{M},\Omega,H)$, where $(\mathcal{M},\Omega)$ is a symplectic manifold and $H\in C^{\infty}(\mathcal{M})$ is a hamiltonian function.

A bijective map $f:\mathcal{M}\longrightarrow\mathcal{N}$ of two manifolds $\mathcal{M}$ and $\mathcal{N}$ is called a diffeomorphism if $f$ and its inverse $f^{-1}$ are differentiable.  $f$ is called $C^{r}$-diffeomorphism if $f$ and  $f^{-1}$ are $r$ times continuously differentiable. In case the manifold $\mathcal{M}$ is second-countable and Hausdorff, then  a diffeomorphisms group  $\mathrm{Diff}(\mathcal{M})$  of $\mathcal{M}$ is the group of all  $C^{r}$ diffeomorphisms of $\mathcal{M}$  to itself.
An action of a Lie group $G$ on $\mathcal{M}$ is a group homomorphism
\begin{equation}
\psi:G\longrightarrow \mathrm{Diff}(\mathcal{M})
\end{equation}
defined by $g\longmapsto\psi_{g}$. We also call
\begin{equation}
\mathrm{ev}_{\psi}:\mathcal{M}\times G\longrightarrow \mathcal{M}
\end{equation}
the evaluation map associated with $\psi$ which is defined by $(p,g)\longmapsto \psi_{g}(p)$.
Next let $\mathrm{Sympl}(\mathcal{M},\Omega)$ denotes the group of symplectomorphisms of $(\mathcal{M},\Omega)$. Then,  we call $\psi$ a symplectic action if it is symplectomorphisms,
\begin{equation}
\psi:G\longrightarrow \mathrm{Sympl}(\mathcal{M},\Omega) \subset\mathrm{Diff}(\mathcal{M}).
\end{equation}
Now, we suppose that $\mathfrak{g}$ is Lie algebra of $G$ and  $\mathfrak{g}^{*}$ is its dual vector space. Then, $\psi$ is a hamiltonian action if there exists a map
\begin{equation}
\mu:\mathcal{M}\longrightarrow \mathfrak{g}^{*}
\end{equation}
that satisfy the following conditions: a)  let $\mu^{X}:\mathcal{M}\longrightarrow \mathbb{R}$ for each $X\in \mathfrak{g}$ defined by $\mu^{X}(p)=\langle\mu(p),X\rangle$ be the component of $\mu$ along $X$. Moreover, let $X^{\sharp}$ be the vector field on $\mathcal{M}$ generated by $\{\exp tX : t\in \mathbb{R}\}\subseteq G$. Then $d\mu^{X}=\imath_{X^{\sharp}}\Omega$, that is, the function $\mu^{X}$ is a hamiltonian function for the vector field $X^{\sharp}$; b) $\mu$ is equivariant with respect to $\psi$ and the coadjoint action $\mathrm{Ad}^{*}$ of $G$ on $\mathfrak{g}^{*}$, that is $\mu\circ\psi_{g}=\mathrm{Ad}^{*}\circ\mu$, for all $g\in G$. The map $\mu$ is called a moment map and the vector $(\mathcal{M},\Omega,G, \mu)$ is called a hamiltonian $G$-space.
For example, let $G=\mathbb{T}^{n}$ be  an $n$-dimensional torus with $\mathfrak{g}\simeq \mathbb{R}^{n}$ and $\mathfrak{g}^{*}\simeq\mathbb{R}^{n}$. Then, a moment map
 \begin{equation}
 \mu:\mathcal{M}\longrightarrow \mathbb{R}^{n}
\end{equation}
for an action of $G$ on $(\mathcal{M},\Omega)$ satisfies the following statement. The function $\mu^{X_{i}}$, where $X_{i}$ is a basis of $\mathbb{R}^{n}$, is a hamiltonian function for $X^{\sharp}_{i}$ and is invariant under the action of torus.
Now, suppose that
 \begin{equation}\psi:\mathbb{T}^{n}\longrightarrow \mathrm{Sympl}(\mathcal{M},\Omega)
\end{equation}
is a hamiltonian action  with moment map $\mu:\mathcal{M}\longrightarrow \mathbb{R}^{n}$, then the image of $\mu$ is called the moment polytope which is convex and is also the convex hull of the images of the fixed points of $\psi$.  An action of $G$ on $\mathcal{M}$ is called effective if it is injective as the following map $G\longrightarrow \mathrm{Diff}(\mathcal{M})$. A compact connected symplectic manifold with an effective hamiltonian action of a torus $\mathbb{T}$ of dimension equal to half the dimension of the manifold,  $\dim \mathbb{T}= \frac{1}{2}\dim \mathcal{M}$, and with a choice of a corresponding $\mu$ is called a symplectic toric manifold.

Next, we will discuss the classification of toric manifold based on
Delzant's theorem and symplectic reduction. But first we define the
Delzant polytope $\Delta\subset\mathbb{R}^{n}$ to be a polytope that
satisfies the following properties: a) each edge is of the form
$v+tu_{i}$ for $t\geq0$ and $u_{i}\in\mathbb{Z}^{n}$, that is the
edges meeting at $v$ are rational; b) For each vertex,
$u_{i}\in\mathbb{Z}^{n}$ can be chosen to be a $\mathbb{Z}$ basis of
$\mathbb{Z}^{n}$; c) there are $n$ edges meeting at $v$. Now, based
on Delzant's theorem, toric manifolds are classified by Delzant
polytope. We have the following bijection
 \begin{equation}\{\text{toric manifolds}\}\longrightarrow\{\text{Delzant polytopes}\}
\end{equation}
which is defined by $\{\mathcal{M}^{2n},\Omega,
\mathbb{T}^{n},\mu\}\longmapsto \mu(\mathcal{M})$. Thus based on
Delzant polytope $\Delta$ we can construct
  a symplectic manifold $(\mathcal{M}_{\Delta},\Omega_{\Delta})$, where $\Omega_{\Delta}$ is
  a reduced symplectic form. One also can show that $(\mathcal{M}_{\Delta},\Omega_{\Delta})$
  is a hamiltonian $\mathbb{T}^{n}$-space with a moment map $\mu_{\Delta}$ which has the
  following image $\mu_{\Delta}(\mathcal{M}_{\Delta})=\Delta$ \cite{Silva1}.
  Thus $(\mathcal{M}_{\Delta},\Omega_{\Delta},\mathbb{T}^{n},\mu_{\Delta})$ is
   a toric manifold based on the Delzant polytope $\Delta$.

Now, let $\Delta$ be an $n$-dimensional polytope which is  Delzant but
also it is a lattice polytope. Then  the set of integral points in $\Delta$ can be defined by
\begin{equation}
P=\mathbb{Z}^{n}\cap\Delta=\{\kappa^{(1)},\kappa^{(2)},\ldots,\kappa^{(k)}\},
\end{equation}
where $k=\# P$ is the number of such points. Based on this construction,
 the convex hull of $P$ is the Delzant polytope. The associated variety $\mathcal{X}_{P}$
 is a toric variety for the complex torus $(\mathbb{C}^{n*})$ and can be embedded in
  $\mathbb{P}^{k-1}$, \begin{equation}
i:\mathcal{X}_{P}\longrightarrow\mathbb{P}^{k-1}.
\end{equation}
The toric variety construction and Delzant's construction
   gives equivalent symplectic toric manifolds \cite{Silva2}. Next, we will
    discuss the equivariant embedding of these constructions into the projective spaces.

We also would like to show how we can construct fans from polytopes. For a linear function $f:\mathbb{R}^{n}\longrightarrow \mathbb{R}$ let $P\subset\mathbb{R}^{n}$ be a polytope and $\mathrm{supp}_{P}f$ be the supporting face of $f$ in $P$. Moreover, let $F$ be the face of $P$. Then, the cone associated  to $F$ is the closure of subset $C_{F,P}\subset\mathbb{R}^{n*}$ which is consisted of all linear functions such that $\mathrm{supp}_{P}f=F$. Now, the collection of $\mathcal{F}_{P}$ of $C_{F,P}$ is called the fan of polytope $P$.
A compact toric variety $\mathcal{X}_{P}$ is called projective if there exists an injective morphism
$$\Phi:\mathcal{X}_{P}\longrightarrow\mathbb{P}^{r}$$ of $\mathcal{X}_{P}$
 into some projective space such that $\Phi(\mathcal{X}_{P})$ is Zariski
 closed in $\mathbb{P}^{r}$.
 Now, let the morphism
$\Phi$  be embedding which is induced by the rational map $\varphi:\mathcal{X}_{P}  \longrightarrow  \mathbb{P}^{r}$
defined by $p \mapsto[z^{m_{0}},z^{m_{1}},\ldots,z^{m_{r}}],$ where $z^{m_{l}}(p)=p^{m_{l}}$ in case $p=(p_{1},p_{2},\ldots, p_{n})$. Then, the rational map $\Phi(\mathcal{X}_{P})$ is the set of common solutions of finitely many monomial equations
\begin{equation}
z^{\beta_{0}}_{i_{0}}z^{\beta_{1}}_{i_{1}}\cdots z^{\beta_{s}}_{i_{s}}=z^{\beta_{s+1}}_{i_{s+1}}z^{\beta_{s+2}}_{i_{s+2}}\cdots z^{\beta_{r}}_{i_{r}}
\end{equation}
which satisfy the following relationships
\begin{equation}
  \beta_{0}m_{0}+\beta_{1}m_{1}+\cdots +\beta_{s}m_{s}=\beta_{s+1}m_{s+1}+\beta_{s+2}m_{s+2}+\cdots +\beta_{r}m_{r}
\end{equation}
and
\begin{equation}
  \beta_{0}+\beta_{1}+\cdots +\beta_{s}=\beta_{s+1}+\beta_{s+2}+\cdots +\beta_{r}
,
\end{equation}
for all $\beta_{l}\in \mathcal{Z}_{\geq 0}$ and $l=0,1,\ldots, r$ \cite{Ewald}. This construction of projective toric variety is very important in the next section.

\section{Space of a single quantum state} \label{sec3}
In this section we will investigate the space of a single quantum state based on a toric manifold. We will also in detail discuss a single qubit state. Let the state of a  single quantum be given by
\begin{equation}
\ket{\Psi}=\sum^{n-1}_{x_{1}=0}\alpha_{x_{1}}\ket{x_{1}},
\end{equation}
 with corresponding  Hilbert space $
\mathcal{H}_{\mathcal{Q}}=\mathcal{H}_{\mathcal{Q}_{1}}=\mathbb{C}^{n}
$.
Let  also $(\mathbb{P}^{n-1},\Omega_{FS})$ be a $n-1$-dimensional complex projective space with corresponding Fubini-Study form. Then, the action of $\mathbb{T}^{n-1}$ on $\mathbb{P}^{n-1}$ is given by
\begin{equation}
(e^{i\theta_{1}}, e^{i\theta_{2}}, \ldots,e^{i\theta_{n-1}})\cdot[\alpha_{0}:\alpha_{1}:\cdots:\alpha_{n-1}]=
[\alpha_{0}:e^{i\theta_{1}}\alpha_{1}:\cdots:e^{i\theta_{n-1}}\alpha_{n-1}]
\end{equation}
and has moment map
\begin{equation}
\mu[\alpha_{0}:\alpha_{1}:\cdots:\alpha_{n-1}]=
-\frac{1}{2}\left(\frac{|\alpha_{1}|^{2}}{|\alpha_{0}|^{2}+\cdots
+|\alpha_{n-1}|^{2}},\ldots,
\frac{|\alpha_{n-1}|^{2}}{|\alpha_{0}|^{2}+\cdots+|\alpha_{n-1}|^{2}}\right).
\end{equation}
Moreover, the fixed points have the following maps
$$
\begin{array}{ccc}
  [1:0:\cdots:0] ~~& \longmapsto & (0,0,\ldots,0) \\
\end{array}
$$
$$
\begin{array}{ccc}
  [0:1:\cdots:0] & \longmapsto & (-\frac{1}{2},0,\ldots,0) \\
  &\vdots&
\end{array}
$$
$$
\begin{array}{ccc}
  [0:0:\cdots:1] & \longmapsto & (0,0,\ldots,-\frac{1}{2}) \\
\end{array}
$$
Let $\Omega=\frac{i}{2}\sum_{k}d\alpha_{k}\wedge d\overline{\alpha}_{k}=\sum_{k}r_{k}dr_{k}\wedge d\theta_{k}$ be a symplectic form on $\mathbb{C}^{n}$. Then we consider the following action of $S^{1}$ on the pair $(\mathcal{M},\omega)=(\mathbb{C}^{n},\Omega)$: $t\in S^{1}\longmapsto \psi_{t}$, where $\psi_{t}$ defined to be multiplication by $t$. Thus the action $\psi$ is hamiltonian with moment map $\mu:\mathbb{C}^{n}\longrightarrow \mathbb{R}$ that is defined by $\alpha\longmapsto -\frac{\|\alpha\|^{2}}{2}+\frac{1}{2}$. In this case $\mu^{-1}(0)=S^{2n-1}$ and the orbit space of zero level of $\mu$ is
\begin{equation}
\mathcal{M}/G=\mu^{-1}(0)/S^{1}=S^{2n-1}/S^{1}=\mathbb{P}^{n-1}.
\end{equation}
This symplectic reduction gives the projective space $\mathbb{P}^{n-1}$ with Fubini-Study symplectic $\Omega_{\mathrm{red}}=\Omega_{\mathrm{FS}}$.

Next, we will in detail illustrate above construction of symplectic manifold
and symplectic toric manifold by considering first the 2-sphere
$\mathcal{M}=S^{2}$ which is related to the one dimensional complex projective space
$S^{3}/S^{1}=\mathbb{P}^{1}\simeq \mathbb{C}^{2}$, by hopf fibration $S^{3}\longrightarrow S^{2}$
and also is  the space of a single-qubit state
\begin{equation}\ket{\Psi}=\sum^{1}_{
x_{1}=0}\alpha_{x_{1}}\ket{x_{1}}=\alpha_{0}\ket{0}
+\alpha_{1}\ket{1}\in \mathcal{H}_{\mathcal{Q}_{1}}=\mathbb{C}^{2}.
\end{equation}
Let $p$ be a point on $\mathcal{M}=S^{2}$. Then the tangent vectors to $S^{2}$ at $p$ could be identified with vectors orthogonal to $p$. Now, let $(r,s)\in T_{p}S^{2}=\{p\}^{\perp}$. Then a  standard symplectic form on $S^{2}$ is given by $\Omega_{p}(r,s)=\langle p,r\times s\rangle$ which is nondegenerate since $\Omega_{p}(r,s)\neq0$ when $u\neq0$. Moreover, on $(\mathcal{M},\Omega_{S})=(S^{2}, d\theta\wedge dh)$ the vector field $X=\frac{\partial}{\partial \theta}$ is hamiltonian function  given by
$dh=\imath_{X}(d\theta\wedge dh)$. As symplectic manifold $S^{1}$ acts on $(\mathcal{M},\Omega_{S})=(S^{2}, d\theta\wedge dh)$ by $\exp(it)\cdot (\theta,h)=(\theta+t,h)$ with moment map $\mu=h$ and moment polytope $[-1,1]$. This construction is also gives moment map for the one dimensional complex projective space
$\mathbb{P}^{1}$ with the Fubini-Study form $\Omega_{FS}=\frac{1}{4}\Omega_{S}$. The action of $S^{1}$ on $\mathbb{P}^{1}$ is given by $\exp(i\theta)\cdot[\alpha_{0}:\alpha_{1}]=[\alpha_{0}:e^{i\theta}\alpha_{1}]$.
The hamiltonian for $\mathbb{P}^{1}$ has moment map
\begin{equation}
\mu[\alpha_{0}:\alpha_{1}]=-\frac{1}{2}\frac{|\alpha_{1}|^{2}}{|\alpha_{0}|^{2}+|\alpha_{1}|^{2}}
\end{equation}
and moment polytope $[-\frac{1}{2},0]$. This construction of moment map for  $\mathbb{P}^{1}$ can
be generalized into multipartite systems.

\section{Bipartite and multipartite quantum states} \label{sec4}
After discussing the single quantum states, now we will
 investigate the structures of multi-qubit quantum systems based on toric manifolds.
First we consider the space of two-qubit states
\begin{equation}\ket{\Psi}=\sum^{1}_{x_{2},
x_{1}=0}\alpha_{x_{2}x_{1}}\ket{x_{2}
x_{1}}=\alpha_{00}\ket{00}+\alpha_{01}\ket{01}+\alpha_{10}\ket{10}+
\alpha_{11}\ket{11}\in\mathcal{H}_{\mathcal{Q}}.
\end{equation}
Now, the action of $\mathrm{T}^{2}$ on $\mathbb{P}^{1}\times\mathbb{P}^{1}$ is given by
\begin{equation}
(\exp(i\theta_{1}),\exp(i\theta_{2}))\cdot([\alpha^{1}_{0}:\alpha^{1}_{1}],
[\alpha^{2}_{0}:\alpha^{2}_{1}])
=([\alpha^{1}_{0}:e^{i\theta_{1}}\alpha^{1}_{1}],[\alpha^{2}_{0}:e^{i\theta_{2}}\alpha^{2}_{1}]).
\end{equation}
The hamiltonian for $\mathbb{P}^{1}\times\mathbb{P}^{1}$ has moment map
\begin{equation}
\mu([\alpha^{1}_{0}:\alpha^{1}_{1}],
[\alpha^{2}_{0}:\alpha^{2}_{1}])=-\frac{1}{2}(\frac{|\alpha^{1}_{1}|^{2}}{|\alpha^{1}_{0}|^{2}
+|\alpha^{1}_{1}|^{2}},\frac{|\alpha^{2}_{1}|^{2}}{|\alpha^{2}_{0}|^{2}
+|\alpha^{2}_{1}|^{2}})
\end{equation}
and moment polytope $[-\frac{1}{2},0]\times[-\frac{1}{2},0]$. Note that we could
also have the following polytope $[-1,1]\times[-1,1]$.

Next we discuss the multi-qubit states
$\ket{\Psi}=\sum^{1}_{x_{m},x_{m-1},\ldots,
x_{1}=0}\alpha_{x_{m}x_{m-1}\cdots x_{1}}$ $\ket{x_{m}x_{m-1}\cdots
x_{1}},
$ by considering the action of $\mathrm{T}^{m}$ on $\overbrace{\mathbb{P}^{1}\times\mathbb{P}^{1}\times\cdots
\times\mathbb{P}^{1}}^{m}$  define by
\begin{eqnarray}\nonumber
&&(\exp(i\theta_{1}),\exp(i\theta_{2}),\ldots,\exp(i\theta_{m}))
\cdot([\alpha^{1}_{0}:\alpha^{1}_{1}],[\alpha^{2}_{0}:\alpha^{2}_{1}],\ldots,
[\alpha^{m}_{0}:\alpha^{m}_{1}])\\\nonumber&&
=([\alpha^{1}_{0}:e^{i\theta_{1}}\alpha^{1}_{1}],[\alpha^{2}_{0}:e^{i\theta_{2}}\alpha^{2}_{1}],
\ldots,[\alpha^{m}_{0}:e^{i\theta_{m}} \alpha^{m}_{1}])
\end{eqnarray}
for which the  hamiltonian has the moment map
\begin{eqnarray}\nonumber&&
\mu([\alpha^{1}_{0}:\alpha^{1}_{1}],
[\alpha^{2}_{0}:\alpha^{2}_{1}],\ldots,[\alpha^{m}_{0}:\alpha^{m}_{1}])
\\\nonumber&&=-\frac{1}{2}(\frac{|\alpha^{1}_{1}|^{2}}{|\alpha^{1}_{0}|^{2}
+|\alpha^{1}_{1}|^{2}},\frac{|\alpha^{2}_{1}|^{2}}{|\alpha^{2}_{0}|^{2}
+|\alpha^{2}_{1}|^{2}},\ldots,\frac{|\alpha^{m}_{1}|^{2}}{|\alpha^{m}_{0}|^{2}
+|\alpha^{m}_{1}|^{2}}).
\end{eqnarray}
Recently, we have investigated the geometrical structure multipartite quantum systems based on algebraic toric varieties. Now, based on our discussion about Delzant polytope and also relation between poltype and fan we  are able to discuss the entanglement properties of multipartite states by embedding the toric manifold $\mathcal{X}_{A}$ in to a projective space.

Let $M=\mathbb{Z}^{m}$ and consider the $m$ cube $P$ centered at the origin
with vertices $(\pm1,\ldots,\pm1)$. This gives the toric variety $\mathcal{X}_{P}=
\mathbb{P}^{1}\times\mathbb{P}^{1}\times\cdots\times\mathbb{P}^{1}$.
 Now, consider the following map
$\Phi:\mathcal{X}_{P}\longrightarrow\mathbb{P}^{2^{m-1}}$.
For this map, $\Phi(\mathcal{X}_{P})$ is  a set of the common
solutions of the following monomial equations
\begin{equation}
z^{\beta_{0}}_{i_{0}}z^{\beta_{1}}_{i_{1}}\cdots
z^{\beta_{2^{m-1}-1}}_{i_{2^{m-1}-1}}=z^{\beta_{2^{m-1}}}_{i_{2^{m-1}}}
\cdots z^{\beta_{2^{m}-1}}_{i_{2^{m}-1}}
\end{equation}
that gives  quadratic polynomials $\alpha_{x_{m}x_{m-1}\ldots
x_{1}}\alpha_{y_{m}y_{m-1}\ldots y_{1}} = \alpha_{x_{m}x_{m-1}\ldots
y_{j}\ldots x_{1}}$ $\alpha_{y_{m}y_{m-1} \ldots x_{j}\ldots y_{1}}$ for
all $j=1,2,\ldots,m$ which  coincides with the Segre variety,
where $\alpha_{x_{m}x_{m-1}\ldots
x_{1}}$ are homogenous coordinates on the projective space $\mathbb{P}^{2^{m-1}}$.
Moreover, we have
\begin{equation}
\Phi(\mathcal{X}_{P})=\mathrm{Specm} \mathbb{C}[\alpha_{00\ldots
0},\alpha_{00\ldots 1},\ldots,\alpha_{11\ldots
1}]/\mathcal{I}(\mathcal{A}) ,
\end{equation}
where
\begin{equation}\mathcal{I}(\mathcal{A})=\langle \alpha_{x_{m}x_{m-1}\ldots
x_{1}}\alpha_{y_{m}y_{m-1}\ldots y_{1}} - \alpha_{x_{m}x_{m-1}\ldots
y_{j}\ldots x_{1}}\alpha_{y_{m}y_{m-1} \ldots
x_{j}\ldots x_{1}}\rangle_{\forall j;x_{j},y_{j}=0,1}
\end{equation}
 and $\mathrm{Specm}$ denotes the maximal spectrum of  a polynomial ring.
This toric variety describes the space of separable states in a multi-qubit quantum systems. We
 can also define a measure of entanglement for multi-qubit state based on a modification of this variety \cite{Hosh4}.

As an illustrative example we will in detail discuss a three-qubit state $\ket{\Psi}=\sum^{1}_{x_{3},x_{2},x_{1}=0}
\alpha_{x_{3}x_{2}x_{1}} \ket{x_{3}x_{2}x_{1}}.$ For this  state the separable state is given by the Segre embedding of
$
\mathbb{P}^{1}\times\mathbb{P}^{1}\times\mathbb{P}^{1}$.
Moreover, let $M=\mathbb{Z}^{3}$ and consider the polytope $P$ centered at the origin with vertices $(\pm1,\pm1,\pm1)$ which is a cube. This gives the toric variety $\mathcal{X}_{P}=\mathbb{P}^{1}\times\mathbb{P}^{1}\times\mathbb{P}^{1}$.  Now, we have the following map
$\Phi:\mathcal{X}_{P}\longrightarrow\mathbb{P}^{7}$.
 In this special case,
$\Phi(\mathcal{X}_{P})$ is the set of common solutions of
finitely many monomial equations
\begin{equation}
z^{\beta_{0}}_{i_{0}}z^{\beta_{1}}_{i_{1}}z^{\beta_{2}}_{i_{2}}z^{\beta_{3}}_{i_{3}}
=z^{\beta_{4}}_{i_{4}}z^{\beta_{5}}_{i_{5}}z^{\beta_{6}}_{i_{6}}z^{\beta_{7}}_{i_{7}}
\Longrightarrow\alpha_{x_{3}x_{2}x_{1}}\alpha_{y_{3}y_{2}y_{1}}
=\alpha_{x_{3}y_{j}x_{1}}\alpha_{y_{3}x_{j}y_{1}},
\end{equation}
where e.g., $\beta_{0}=\beta_{1}=\beta_{4}=\beta_{5}=1$,
$\beta_{2}=\beta_{3}=\beta_{6}=\beta_{7}=0$, and $j=1,2,3$. Then,
the projective toric variety is gives                                                                                                                                            by
\begin{equation}
\Phi(\mathcal{X}_{P})=\mathrm{Specm}
\mathbb{C}[\alpha_{000},\alpha_{001},\ldots,\alpha_{111}]/
\mathcal{I}(\mathcal{A}),
\end{equation}
where $\mathcal{I}(\mathcal{A})
=\langle \alpha_{x_{3}x_{2}x_{1}}\alpha_{y_{3}y_{2}y_{1}} - \alpha_{x_{3}
y_{j} x_{1}}\alpha_{y_{3}
x_{j}y_{1}}\rangle_{\forall j=1,2,3;x_{j},y_{j}=1,2}$.
One can also find the following relation between this construction and $3$-tangle $\tau$ and hyperdeterminat $D_{3}$ which is given by
\begin{eqnarray}\label{HH}
&&\tau/4=D_{3}= d_{1}-2d_{2}+4d_{4}
\end{eqnarray}where
$d_{1}=\alpha^{2}_{000}\alpha^{2}_{111}
+\alpha^{2}_{001}\alpha^{2}_{110}+\alpha^{2}_{010}\alpha^{2}_{101}+
\alpha^{2}_{100}\alpha^{2}_{011}$,
$d_{2}=\alpha_{000}\alpha_{001}\alpha_{110}\alpha_{111}
+\alpha_{000}\alpha_{010}\alpha_{101}\alpha_{111}+
\alpha_{000}\alpha_{100}\alpha_{011}\alpha_{111}
+\alpha_{001}\alpha_{010}\alpha_{101}\alpha_{110}+
\alpha_{001}\alpha_{100}\alpha_{011}\alpha_{110}
+\alpha_{010}\alpha_{100}\alpha_{100}\alpha_{101}$, and
$d_{4}=\alpha_{000}\alpha_{110}\alpha_{101}\alpha_{011}
+\alpha_{111}\alpha_{100}\alpha_{010}\alpha_{001}$
which is a good measure of entanglement for three-qubit systems \cite{Coffman}.
 In this expression $d_{1}$ are diagonal lines in the three cube (which we have shown to be
 the toric variety $\mathcal{X}_{P}$), $d_{2}$ are the diagonal planes, and $d_{4}$ are tetrahedrons.

Now, we will review the construction of concurrence, $m$-tangle, polynomial, and geometrical invariants. We also
discuss how these measures of entanglement are related to symplectic toric variety.
First, we
introduce a complex conjugation operator $\mathcal{C}_{m}$ that
acts on the multipartite quantum state $\ket{\Psi}$  as
\begin{equation}\label{cong}
\ket{\Psi^{*}}=\mathcal{C}_{m}\ket{\Psi}=\sum^{1}_{x_{m},x_{m-1},\ldots,
x_{1}=0}\alpha^{*}_{x_{m}x_{m-1}\cdots x_{1}}\ket{x_{m}x_{m-1}\cdots x_{1}}
.
\end{equation}
Then, the concurrence of two-qubit states is defined as
 \begin{equation}\mathcal{C}(\Psi)=|\langle\Psi\ket{\widetilde{\Psi}}|^{2},
 \end{equation}
  where  the tilde represents the "spin-flip" operation
 $\ket{\widetilde{\Psi}}=\sigma_{y}\otimes
  \sigma_{y}\ket{\Psi^{*}}$
,  $\ket{\Psi^{*}}$ is defined by equation (\ref{cong}), and $\sigma_{y}=\left(%
\begin{array}{cc}
  0 & -i \\
  i & 0 \\
\end{array}%
\right)$ is a Pauli spin-flip operator
\cite{Wootters98,Wootters00}. This construction can be generalized to a multi-qubit system by defining
\begin{equation}\ket{\widetilde{\Psi}}=\sigma^{\otimes m}_{y}
\ket{\Psi^{*}},
\end{equation}
where $\sigma^{\otimes m}_{y}$ denotes $m$-folds tensor product of $\sigma_{y}$. Next, we define $m$-tangle as
 \begin{equation}\tau_{m}=|\langle\Psi\ket{\widetilde{\Psi}}|^{2}
 \end{equation}
for every even $m$-qubit system \cite{Wong01}.
Now, we in detail discuss a four-qubit state which is given by
\begin{equation}\label{Mstate}
\ket{\Psi}=\sum^{1}_{x_{4}=0}\sum^{1}_{x_{3}=0}\sum^{1}_{x_{2}=0}\sum^{1}_{x_{1}=0}
\alpha_{x_{4}x_{3}x_{2} x_{1}} \ket{x_{4}x_{3}x_{2} x_{1}}\in\mathcal{H}_{\mathcal{Q}},
\end{equation}
where
$\mathcal{H}_{\mathcal{Q}}=\mathcal{H}_{\mathcal{Q}_{1}}\otimes
\mathcal{H}_{\mathcal{Q}_{2}}\otimes\mathcal{H}_{\mathcal{Q}_{3}}\otimes\mathcal{H}_{\mathcal{Q}_{4}}=
\mathbb{C}^{2}\otimes\mathbb{C}^{2}\otimes\mathbb{C}^{2}\otimes\mathbb{C}^{2}$ and $x=x_{4}2^{3}+
x_{3}2^{2}+x_{2}2^{1}+x_{1}2^{0}$. The first polynomial invariant $H$ of degree 2 is defined by
\begin{eqnarray}
\nonumber
  H &=&\alpha_{0}\alpha_{15}-\alpha_{1}\alpha_{14}-\alpha_{2}\alpha_{13}+
         \alpha_{3}\alpha_{12}-\alpha_{4}\alpha_{11}+\alpha_{5}\alpha_{10}+
         \alpha_{6}\alpha_{9}-
         \alpha_{7}\alpha_{8}
\end{eqnarray}
and is one of hyperdeterminants introduced by Cayley \cite{Thibon}. Next, we will review the construction of geometrical four-qubit invariants presented in \cite{Levay}. Using the notation which we have already introduced, we define the following four column vectors,
\begin{equation}
\mathcal{A}\equiv\left(
               \begin{array}{c}
                 \alpha_{0} \\
                 \alpha_{1}\\
                 \alpha_{2} \\
                 \alpha_{3} \\
               \end{array}
             \right)~~\mathcal{B}\equiv\left(
               \begin{array}{c}
                 \alpha_{4} \\
                 \alpha_{5}\\
                 \alpha_{6} \\
                 \alpha_{7} \\
               \end{array}
             \right)~~\mathcal{C}\equiv\left(
               \begin{array}{c}
                 \alpha_{8} \\
                 \alpha_{9}\\
                 \alpha_{10} \\
                 \alpha_{11} \\
               \end{array}
             \right)~~\mathcal{D}\equiv\left(
               \begin{array}{c}
                 \alpha_{12} \\
                 \alpha_{13}\\
                 \alpha_{14} \\
                 \alpha_{15} \\
               \end{array}
             \right),
\end{equation}
where $\mathcal{A},\mathcal{B},\mathcal{C},\mathcal{D}\in\mathbb{C}^{4}$. Moreover, let
$g:\mathbb{C}^{4}\times \mathbb{C}^{4}\longrightarrow\mathbb{C}$ be a bilinear form such that
\begin{equation}
(\mathcal{A},\mathcal{B})\mapsto g (\mathcal{A},\mathcal{B})
\equiv\mathcal{A}\cdot\mathcal{B}=g_{\alpha\beta}\mathcal{A}^{\alpha}\mathcal{B}^{\beta}
=\mathcal{A}_{\alpha}\mathcal{B}^{\alpha},
\end{equation}
where $\alpha,\beta=0,1,2,3$, $g=\mathcal{J}\otimes \mathcal{J} $ and $\mathcal{J}=\left(
           \begin{array}{cc}
             0 & 1 \\
             -1 & 0 \\
           \end{array}
         \right)$ which satisfy $\mathcal{J}^{2}=-\mathbb{I}$ and $\mathcal{M}\mathcal{J}\mathcal{M}^{T}=\mathcal{J}$ for $\mathcal{M}\in SL(2,\mathbb{C})$. Then, the first stochastic local operations and classical communication (SLOCC) invariant for four-qubit states are then given by
         \begin{equation}
         I_{1}=\frac{1}{2}(\mathcal{A}\cdot\mathcal{D}-\mathcal{B}\cdot\mathcal{C}).
         \end{equation}
In this case the four-tangle can also be expressed in terms of polynomial and geometrical invariants as follows
\begin{eqnarray}\nonumber
\tau_{4}&=&2|\alpha_{\kappa_{4}\kappa_{3}\kappa_{2}\kappa_{1}}
\alpha_{\lambda_{4}\lambda_{3}\lambda_{2}\lambda_{1}}
\alpha_{\mu_{4}\mu_{3}\mu_{2}\mu_{1}}
\alpha_{\nu_{4}\nu_{3}\nu_{2}\nu_{1}}\\\nonumber&&
\varepsilon_{\kappa_{4}\lambda_{4}}
\varepsilon_{\kappa_{3}\lambda_{3}}
\varepsilon_{\kappa_{2}\lambda_{2}}
\varepsilon_{\mu_{4}\nu_{4}}
\varepsilon_{\mu_{3}\nu_{3}}
\varepsilon_{\mu_{2}\nu_{2}}
\varepsilon_{\kappa_{1}\mu_{1}}
\varepsilon_{\lambda_{1}\nu_{1}}|
\\\nonumber&=&4|I_{1}|^{2}=|H|^{2}
\end{eqnarray}
where $I_{1}=\frac{1}{2}H$ and $\kappa_{j},\lambda_{j},\mu_{j},\nu_{j}\in\{0,1\}$. Thus the terms in four-tangle are the diagonal lines in  the symplectic toric variety $\mathcal{X}_{P}$ of four qubit systems, where the polytope $P$ centered at the origin with vertices $(\pm1,\pm1,\pm1,\pm1)$ is a four cube. This construction can also be generalized into $m$-qubit systems whenever $m$ is an even number.

In summary we have studied the symplectic structures of quantum systems based on symplectic toric manifolds and their moment maps. Using the Delzant's construction we were also able to establish relations between symplectic toric manifolds, algebraic toric varieties and entanglement properties of bipartite and multipartite  quantum systems. The construction of moment maps and toric varieties give useful combinatorial information about complex projective spaces of multipartite quantum systems that enable us to  investigate the entanglement properties of multipartite quantum systems. These advantages of the moment map and toric variety make them good candidates for further consideration that possibly could lead to new  applications of these mathematical structures in the field of quantum information processing.

\begin{flushleft}
\textbf{Acknowledgments:}  The  work was supported  by the Swedish Research Council (VR).
\end{flushleft}


\begin{thebibliography}{22}
 \bibitem{Miyake} A. Miyake and M. Wadati Quant. Inf.
Comp. 2 (Special), 540-555 (2002).
\bibitem{Briand2}E. Briand, J.-G. Luque, J.-Y. Thibon,
and F. Verstraete, J. Math. Phys. {\bf 45} (2004) 4855,
  quant-ph/0306122.
 \bibitem{Levay1} P. L\'{e}vay, J. Phys. A: Math. Gen. {\bf 38} (2005) 9075-9085.
 \bibitem{Hosh4} H. Heydari, J. Math. Phys. {\bf47}, 012103 (2006).
 \bibitem{Vafa} K. Hori, S. Katz, A. Klemm, R. Pandharipande, R. Thomas, C. Vafa, R. Vakil, and E. Zaslow {\it Mirror Symmetry}, AMS, (2003).
\bibitem{Hosh1} H. Heydari, {\it Mathematical Physics Research Developments}, Edited by M. B. Levy, Nova publisher, (2009) pp. 589-604.
\bibitem{Hosh2} H. Heydari, e-print arXiv:1001.3245v1 [quant-ph].
 \bibitem{Guillemin1} V. Guillemin and S. Sternberg, {\it Symplectic techniques in physics}, CUP, 1984.
  \bibitem{Audin}   M. Audin, {\it Torus Actions On Symplectic Manifolds}, Progress In Mathematics, 2Ed, Birkhauser, (2003).
 \bibitem{Mcduff}    D.  Mcduff  and D. Salamon, {\it Introduction To Symplectic Topology}, Oxford, 1998.
  \bibitem{Guillemin2} V. Guillemin, {\it Moment maps and combinatorial invariants of hamiltonian $t^{n}$-spaces}, Birkhauser, (1994).
\bibitem{Silva1} A. Cannas da Silva,
 {\it Lectures on Symplectic Geometry},  Springer LNM $\sharp$1764, (2000).
 \bibitem{Silva2} M. Audin, A. Cannas da Silva, and E. Lerman, {\it Symplectic Geometry of Integrable Hamiltonian Systems},
 (Advanced Courses in Mathematics - CRM Barcelona), Birkhauser, Berlin (2003).
\bibitem{Griff78} P. Griffiths and J. Harris, {\it Principles of
  algebraic geometry}, Wiley and Sons, New York, 1978.
  \bibitem{Mum76} D. Mumford, {\it Algebraic Geometry I,
Complex Projective Varieties}, Springer-Verlag, Berlin, (1976).
 \bibitem{Hart77} R. Hartshorne, {\it Algebraic Geometry}, Springer-Verlag, New York, (1977).

  \bibitem{Ewald}G. Ewald, {\it Combinatorial Convexity and Algebraic Geometry}, Springer, 1995.
   \bibitem{Coffman} V. Coffman, J. Kundu, and W. K. Wootters, Phys. Rev. A {\bf 61}, 052306 (2000).
       \bibitem{Wootters98} W. K. Wootters,  Phys. Rev. Lett. {\bf 80}, 2245 (1998).
\bibitem{Wootters00} W. K. Wootters, Quantum Information and Computaion, Vol. 1, No. 1 (2000) 27-44, Rinton Press.
\bibitem{Wong01} A. Wong and N. Christensen Phys. Rev. A {\bf 63}, 044301 (2001).
\bibitem{Jaeger07} G. Jaeger, {\it Quantum Information  An Overview }, Springer, New York 2007.
\bibitem{Thibon} J.-G. Luque and J.-Y. Thibon, J. Math. Phys. {\bf 45} (2004) 4855.
      \bibitem{Levay}P. Levay, J. Phys. A: Math. Gen. {\bf 39} (2006) 9533-9545.
\end{thebibliography}
\end{document}